\documentclass[final,authoryear,5p]{elsarticle}

\usepackage{epsfig}

\usepackage{amssymb}

\usepackage[ps2pdf,%
a4paper=true,%
breaklinks=true,%
colorlinks=true,%
pdfauthor={G.\ Rauw \& E.\ Mossoux},%
pdftitle={Fe xxv line profiles in colliding wind binaries}%
]{hyperref}

\journal{New Astronomy}

\begin{document}

\begin{frontmatter}



\title{Fe {\sc xxv} line profiles in colliding wind binaries}


\author{Gregor Rauw\corref{cor}}
\address{Groupe d'Astrophysique des Hautes Energies, Institut d'Astrophysique et de G\'eophysique, Universit\'e de Li\`ege, All\'ee du 6 Ao\^ut, 19c, B\^at B5c, 4000 Li\`ege, Belgium}
\cortext[cor]{Corresponding author}
\ead{rauw@astro.ulg.ac.be}

\author{Enmanuelle Mossoux}
\address{Observatoire Astronomique de Strasbourg, Universit\'e de Strasbourg, CNRS, UMR 7550, 11 rue de l'Universit\'e, 67000 Strasbourg, France}
\ead{enmanuelle.mossoux@astro.unistra.fr}

\author{Ya\"el Naz\'e\fnref{footnote4}}
\address{Groupe d'Astrophysique des Hautes Energies, Institut d'Astrophysique et de G\'eophysique, Universit\'e de Li\`ege, All\'ee du 6 Ao\^ut, 19c, B\^at B5c, 4000 Li\`ege, Belgium}
\fntext[footnote4]{Research Associate FRS-FNRS (Belgium)}
\ead{naze@astro.ulg.ac.be}

\begin{abstract}
Strong wind-wind collisions in massive binaries generate a very hot plasma that frequently produces a moderately strong iron line. The morphology of this line depends upon the properties of the wind interaction zone and its orientation with respect to the line of sight. As the binary components revolve around their common centre of mass, the line profiles are thus expected to vary. With the advent of the next generation of X-ray observatories ({\it Astro-H}, {\it Athena}) that will offer high-resolution spectroscopy above 6\,keV, it will become possible to exploit these changes as the most sensitive probe of the inner parts of the colliding wind interaction. Using a simple prescription of the wind-wind interaction in an early-type binary, we have generated synthetic line profiles for a number of configurations and orbital phases. These profiles can help constrain the properties of the stellar winds in such binary systems.
\end{abstract}


\end{frontmatter}

\parindent=0.5 cm

\section{Introduction}
{\it Chandra} and {\it XMM-Newton} have opened up the era of high-resolution X-ray spectroscopy for moderately bright X-ray sources. In the case of massive stars, HETG and RGS spectra unveiled for the first time some details of the morphology of the spectral lines \citep[see e.g.][]{GN}. For presumably single massive stars, where the X-ray emission arises from hot plasma embedded in the stellar wind, a number of observational and theoretical studies were undertaken with the goal to connect the line morphologies to the properties of the stellar wind \citep[e.g.][]{Feldmeier,OC2,Oskinova,Cohen,Herve,lamCep}. 

In massive binary systems, the X-ray emission not only arises from the individual winds of both stars, but also from shock-heated plasma inside the wind interaction zone between the stars \cite[e.g.][]{SBP}. At the shock front between the winds, the kinetic energy normal to the shock is converted into heat, thereby generating a very hot plasma. Since the plasma temperature exceeds the values typically reached in the shocks that prevail in the winds of most single massive stars, the detection of such a hot plasma provides strong hints of a wind-wind interaction. The most obvious signature of such a high-temperature plasma is the moderately strong iron line near 6.7\,keV, commonly called the Fe K line, that can be observed in some systems. Whilst numerous studies have been devoted to colliding winds over the last two decades \citep[for a general review on colliding winds see][]{RN}, most of them focused on broadband medium-resolution spectroscopy. Based on density and velocity maps obtained from hydrodynamical simulations, \citet{Henley} presented calculations of theoretical X-ray line profiles of colliding wind binaries and of their variations along the orbit for a range of orbital and wind parameters. Their study focused on the Ly$\alpha$ transitions of O {\sc viii}, Ne {\sc x}, Mg {\sc xii}, Si {\sc xiv} and S {\sc xvi}. These lines are located at energies below 3\,keV, in the sensitivity band of the current high-resolution spectrographs. However, most of these lines are also emitted by the plasma in the winds of each individual binary component outside the wind-wind interaction zone. Their profiles are thus often contaminated, or even dominated, by contributions from the intrinsic emissions of the stars. Moreover, as shown by \citet{Henley}, photoelectric absorption by the cool unshocked winds affects the morphology of the lines at longer wavelengths, especially in the case of the O, Ne, and Mg L$\alpha$ lines. Both effects render the interpretation of the observed line profiles in terms of colliding winds rather difficult. 

The Fe K feature offers a promising way out of this dilemma. Indeed, except for some stars featuring a strong magnetic field, this line is not seen in the X-ray spectra of single massive stars. Furthermore, at energies near 6.7 -- 7.0\,keV, it is essentially unaffected by absorption in the unshocked winds. Moreover, it is rather isolated, thereby limiting the problems of blends with other species. Unfortunately though, the iron line falls outside the sensitivity range of the current generation of high-resolution spectrographs. However, this situation is about to change in the near future thanks to the JAXA mission {\it Astro-H} \citep{AstroH} and, in the more distant future, thanks to the ESA observatory {\it Athena} \citep{Nandra}. Each of these missions will carry a high-resolution bolometric spectrograph optimized for a high-sensitivity and high-resolution coverage of the spectral region around 6.0 -- 7.0\,keV. {\it Astro-H} will be equiped with the {\it Soft X-ray Spectrometer} \citep[SXS,][]{SXS}, whilst {\it Athena} will host the {\it X-ray Integral Field Unit} \citep[XIFU,][]{XIFU,Ravera}. Whilst these instruments are primarily designed to address other scientific questions, they can also be efficiently used for the purpose of the study of colliding wind binaries \citep{SR}. 

In this context, we address here the question of the morphology of the Fe line at 6.7\,keV and its variations with orbital phase. 

\section{The iron line in the spectra of colliding wind binaries}
Before we turn to the discussion of our models, we need to briefly consider the nature of the Fe K line that is observed in the spectra of colliding wind binaries at energies around 6.7\,keV. As shown by spectral fits, this line is compatible with emission from highly ionized iron in a thermal plasma of high temperature $T$. The emissivity of the triplet of helium-like Fe\,{\sc xxv} between 6.64 and 6.70\,keV peaks near $\log{T} \sim 7.8$ (i.e.\ 63\,MK or $kT = 5.4$\,keV). Conversely, the emissivity of the Ly$\alpha$ doublet of hydrogen-like Fe\,{\sc xxvi} peaks at $\log{T} \sim 8.2$ (i.e.\ 158\,MK or $kT = 13.7$\,keV). Considering that typical wind velocities of massive stars range between 1000 and 3000\,km\,s$^{-1}$, the Rankine-Hugoniot condition for strong shocks implies typical post-shock temperatures between 1.2 and 10.5\,keV. This leads us to conclude that the iron line of the majority of the colliding wind binaries should be associated with the Fe\,{\sc xxv} helium-like triplet. 

Helium-like triplets consist of a resonance line, an intercombination doublet and a forbidden line. In the case of Fe {\sc xxv}, the energies of these lines, taken from the AtomDB database \citep{AtomDB}, are 6.7004\,keV (resonance line), 6.6823 and 6.6676\,keV (intercombination doublet), and 6.6366\,keV (forbidden line). 

\begin{figure}
\begin{center}
\includegraphics*[width=0.45\textwidth,angle=0]{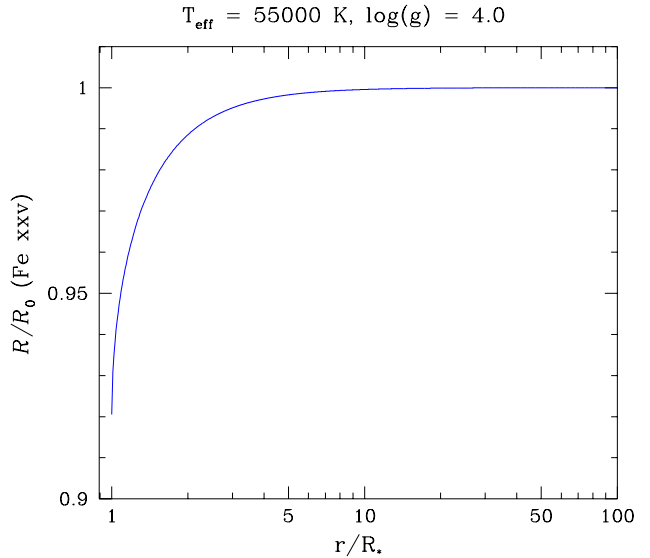}
\end{center}
\caption{Modification of the ${\cal R}$ ratio between the forbidden and intercombination lines in the Fe\,{\sc xxv} triplet formed in the wind of an O-type star with $T_{\rm eff} = 55\,000$\,K and $\log{g} = 4.0$. \label{figure1}}
\end{figure}

As shown e.g.\ by \citet{Blumenthal} and \citet{Porquet}, the forbidden line of the helium-like triplets of light elements can be strongly suppressed as a result of either collisional excitation or radiative pumping of the electrons from its upper level (2\,$^3S_1$) to the upper levels of the intercombination doublet (2\,$^3P_{1,2}$). The latter effect has been used extensively to study the location of the X-ray emitting plasma in the winds of massive stars \citep[e.g.][]{Leutenegger06}. Does a similar effect also exist for the Fe\,{\sc xxv} triplet? To address this question, we follow the formalism of \citet{Blumenthal,Porquet} and \citet{Leutenegger06}. These authors have shown that the ratio ${\cal R}$ between the forbidden and intercombination lines can be written
\begin{equation}
    \label{eq:1}
{\cal R} = \frac{{\cal R}_0}{1 + 2\,w(r)\,\frac{\psi_*}{\psi_c} + \frac{N_e}{N_c}}
  \end{equation}
Here ${\cal R}_0$ is the ratio between the forbidden and intercombination line in the absence of photospheric light and collisional excitation. $N_e$ is the local electron density and $N_c$ is the critical electron density above which collisional excitation from 2\,$^3S_1$ to 2\,$^3P_{1,2}$ becomes important. The photoexcitation rate $\psi_*$ due to the photospheric radiation field is multiplied by the dilution factor 
\begin{equation}
w(r) = \frac{1}{2}\,\left(1 - \sqrt{1 - \frac{R_*^2}{r^2}}\right)
\end{equation}
at a distance $r$ from the centre of the star of radius $R_*$. This value is compared to a critical rate $\psi_c$. In the case of the Fe\,{\sc xxv} triplet, \citet{Blumenthal} quote values of $N_c = 4.7 \times 10^{16}$\,cm$^{-3}$ and $\psi_c = 1.23 \times 10^8$\,s$^{-1}$. 

For a solar composition, $N_c$ corresponds to $\rho_c = 2.3 \times 10^{-8}$\,g\,cm$^{-3}$. This value is several orders above the typical density in adiabatic post-shock regions \citep[e.g.][]{Pittard09}. The suppression of the forbidden line by collisional excitation can thus be neglected. 

Concerning the impact of photoexcitation, the relevant transitions that will pump the electrons are located in the EUV at wavelengths of 271.4 and 396.7\,\AA\ \citep{NIST}. We use the fluxes at the stellar surface predicted by the non-LTE plane-parallel TLUSTY stellar atmosphere model and available via the Ostar2002 grid \citep{Ostar2002} to estimate $\psi_*$. If we consider a very hot main-sequence star with $T_{\rm eff} = 55\,000$\,K, we find that the impact of photospheric radiation on the Fe\,{\sc xxv} triplet is rather marginal (of order 8\% at the stellar surface, see Fig.\,\ref{figure1}) and drops rapidly when we move out into the stellar wind, at radii $\geq 5\,R_*$, where the wind-wind collision occurs. The main reason for this weak effect is of course the high value of $\psi_c$: even such a hot star does not provide enough radiation at the relevant wavelengths to produce a strong pumping. We can thus conclude that we can safely neglect the impact of photospheric radiation on the relative strengths of the components of the Fe\,{\sc xxv} triplet.
  
\section{A simplified model for the wind-wind interaction}  
Depending on the efficiency of radiative cooling in the post-shock plasma, colliding wind binaries are frequently divided into two broad categories \citep{SBP}. 

The plasma is said to be in the adiabatic regime when the post-shock densities are sufficiently low for the cooling time ($t_{\rm cool}$) to be much longer than the escape time ($t_{\rm esc}$) from the shock region. The efficiency of radiative cooling is thus expressed via the parameter $\chi = t_{\rm cool}/t_{\rm esc}$. This situation mainly occurs in wide, long-period binaries and/or systems consisting of stars with relatively low mass-loss rates. The corresponding wind interaction zone is quite smooth and can usually be well modelled with hydrodynamical simulations. The X-ray emission should scale as $\dot{M}^2\,v^{-3.2}\,d^{-1}$ where $\dot{M}$, $v$ and $d$ are the mass-loss rate, the pre-shock wind velocity and the separation between the stars, respectively.

Conversely, if the post-shock plasma density is high, radiative cooling happens very quickly and the wind interaction zone is then said to be radiative. This concerns close binary systems hosting stars with large mass-loss rates, but can also occur in wide, eccentric systems around periastron passage. From first principles, the X-ray emission of these systems should scale with the kinetic power of the incoming stellar winds. However, the observed level of X-ray emission is usually much lower than theoretically expected. This could be due to radiative inhibition \citep{SP}, radiative braking \citep{Gayley} or the strong hydrodynamical instabilities that strongly distort the wind interaction zone \citep{Kee}. 

In our present work, we consider mainly systems that can be described as undergoing an adiabatic wind interaction. We further assume that the winds collide at their terminal velocity $v_{\infty}$, thereby neglecting wind acceleration, radiative inhibition and braking. Since the emission of interest here arises in the inner parts of the interaction zone, we neglect the impact of the orbital motion and the shocks are considered to be axisymmetric about the binary axis. We then use the analytical solution of \citet{Canto} to describe the properties of the wind interaction zone. In principle, the formalism of \citet{Canto} holds for situations where the shocked plasma forms a thin shell around the contact discontinuity, which is the case if both winds are highly radiative. Although we consider here mainly situations where the shocks are adiabatic, the position of the contact discontinuity is unlikely to drastically differ from the predictions of \citet{Canto}, as was shown by \citet{PP08} and \citet{Pittard09}.

\begin{figure}
\begin{center}
\includegraphics*[width=0.45\textwidth,angle=0]{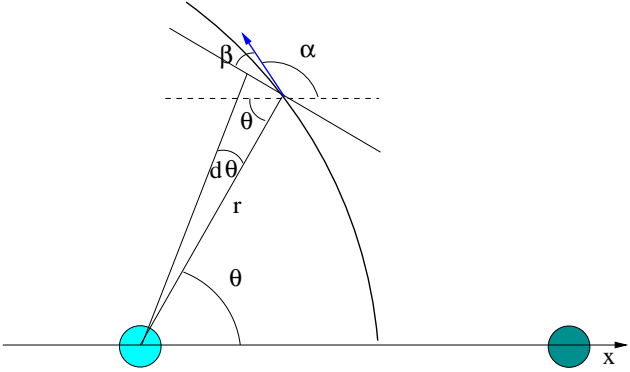}
\end{center}
\caption{Geometry of our model of the wind interaction zone. The star with the weaker wind is on the left. The contact discontinuity between the winds is shown by the thick black line. The blue vector indicates the tangential velocity at the point of coordinates $(r, \theta)$. $\beta = \pi/2 - \alpha + \theta$ is the angle between the tangential velocity and the normal to the position vector. \label{figure2}}
\end{figure}

For a given value of the wind momentum ratio 
\begin{equation}
\eta = \frac{\dot{M_2}\,v_{\infty,2}}{\dot{M_1}\,v_{\infty,1}}
\end{equation} 
solving equation (28) of \citet{Canto} by means of a Newton-Raphson scheme allows us to estimate the asymptotic opening angle of the shock $\theta_{\infty}$ as seen from the star with the weaker wind. The interval $[0,\theta_{\infty}]$ is then discretized into 200 steps $\theta_k$. For each angle, equations (23) and (24) of \citet{Canto} are solved to compute the shape of the contact discontinuity, i.e.\ to establish the relation between $r$ and $\theta$ (see Fig.\,\ref{figure2}). Along with the adopted mass-loss rates and wind velocities, equations (29) and (30) of \citet{Canto} further allow us to estimate the mass surface density $\sigma$ of the wind interaction zone and the tangential velocity along the contact dicontinuity. The resulting shape and velocity vector are then rotated about the binary axis to generate an axisymmetric shock cone, discretized into $200 \times 360$ 2-D cells, along with the associated velocity field. The thickness of these cells is computed as the ratio of the surface density $\sigma$  divided by the post-shock density $\rho_s$. Since we are dealing with an adiabatic wind interaction zone, we use the Rankine-Hugoniot condition for strong shocks $\rho_s = 4\,\rho_w$ where $\rho_w$ is the pre-shock wind density. The volume associated with a cell is then calculated following 
\begin{equation}
    \label{eq:2}
dV = \frac{r^2\,\sin{\theta}\,d\theta\,d\phi}{\sin{(\alpha - \theta)}}\,\frac{\sigma}{\rho_s}
  \end{equation}
(see Fig.\,\ref{figure2}). Here $d\phi = 1^{\circ}$ is the incremental step of the azimuthal angle used to build the axisymmetric shock cone. $\alpha$ is the angle between the tangential velocity and the direction of the binary axis. 

The associated contribution to the line emission is given by
\begin{equation}
    \label{eq:3}
d\epsilon = Q\,n_e\,n_H\,dV
  \end{equation}
where $n_e$ and $n_H$ are the electron and hydrogen number densities in the post-shock region and $Q$ is the line emissivity at the temperature of the post-shock plasma inside the cell assuming ionization equilibrium. The post-shock temperature is computed assuming that the velocity normal to the shock is entirely thermalized \citep[see][]{SBP} and that the electrons and ions have equalized their temperatures. For wind-wind interactions in wide binary systems, these two points, ionization equilibrium and equal electron and ion temperatures ($T_e = T_{\rm ion}$), are clearly approximations. Indeed, as pointed out by several authors \citep[e.g.][]{Usov,ZS,Zhekov,Pollock}, the shocked plasma is probably out of ionization equilibrium and the shocks are likely collisionless, implying $T_e \leq T_{\rm ion}$. However, these approximations are frequently used \citep[including in the work of][]{Henley}. Moreover, they mainly affect the strengths of the simulated lines and should be much less important for their profiles. 

\begin{figure}  
\begin{center}
\includegraphics*[width=0.45\textwidth,angle=0]{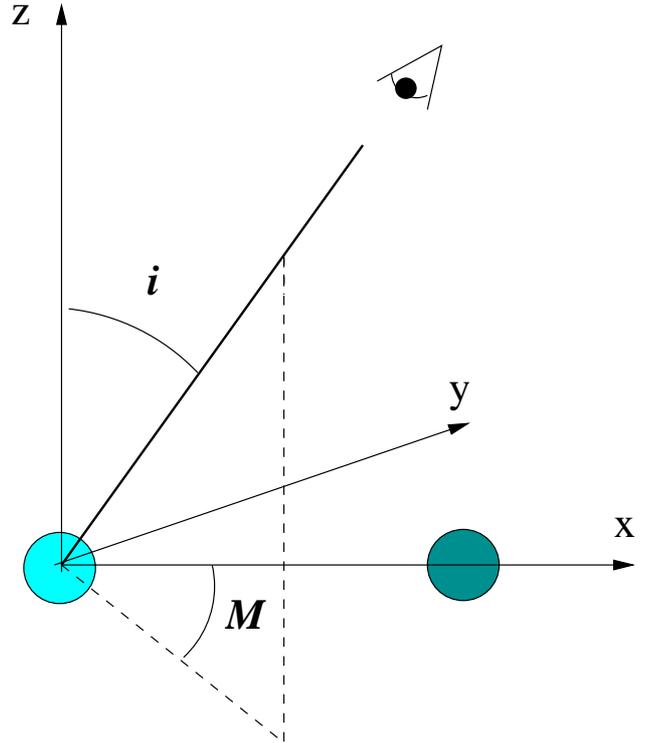}
\end{center}
\caption{Illustration of the angles used to define $\Theta$. The star with the weaker wind is the one at the origin of the axes. The $x$ and $y$ axes are inside the orbital plane and rotate as the stars move around each other on their orbit. \label{figure3}}
\end{figure}

For a given pair of orbital inclination $i$ and orbital phase $\Phi$, we determine the direction of the line-of-sight and project the tangential velocity of each cell onto this direction to obtain the line-of-sight velocity of the gas in the cell. If aberration of the shock due to orbital motion can be neglected, then the orientation of the line-of-sight can be characterized via a single angle $\Theta$ with 
\begin{equation}
\cos(\Theta) = \sin(i)\,\cos(M)
\label{eq:Theta}
\end{equation} 
Here $M = v + \omega -\frac{3\,\pi}{2}$ where $v$ is the true anomaly and $\omega$ is the longitude of periastron of the primary star, which we consider to be the star with the stronger wind (see Fig.\,\ref{figure3}).
 
Although the effect of photoelectric absorption by the winds is very low at the energies of interest here, we have nevertheless accounted for the optical depth along the line-of-sight. Using the ionized wind opacity model of \citet{HD108}, we estimate the opacity of the wind at 6.7\,keV to be 0.7\,cm$^2$\,g$^{-1}$ for solar abundances. This is about a factor 100 and 40 lower than the equivalent opacity at the energies of the O {\sc viii} L$\alpha$ and Ne {\sc x} L$\alpha$ lines, respectively. To compute the optical depth for a given sightline, we first check whether it intersects any of the stars and whether it crosses the wind of one or two stars. If the cell is occulted by one of the stars, its contribution to the observed profile is set to zero. Otherwise, the optical depth along the sightline is evaluated accounting for the appropriate opacities of each wind that is crossed. 

Thermal broadening is accounted for by distributing the flux of the cell over a Gaussian profile centered on the line-of-sight velocity. Given the high mass of the Fe ions, its effect remains quite moderate though (typically $\sigma_{\rm therm} = 86$\,km\,s$^{-1}$ for a 50\,MK plasma).     

Finally the synthetic line profile is obtained as an histogram of the contributions of all the cells according to their line-of-sight velocity. We assume a constant energy resolution of the instrument of 3\,eV (FWHM) at the relevant energies. The synthetic profiles are computed with line-of-sight velocity steps of 50\,km\,s$^{-1}$, and are then binned onto line-of-sight velocity bins of 150\,km\,s$^{-1}$ corresponding to a bit more than one resolution element.   

\section{Results}
In this section, we present the output of our model for several sets of parameters, including systems with circular or elliptical orbits. As pointed out above, the Fe {\sc xxv} line consists of four components which are closely spaced in wavelength and will thus be blended over parts of the parameter space. To illustrate some effects, we thus start our discussion with the simulation of a simpler case, where we consider a single line. 

\subsection{Systems with circular orbits \label{circular}}
Let us consider early-type binaries with circular orbits. As a first step, we compare the line profiles obtained for three different values of the wind momentum ratio $\eta$ and seen under five different viewing angles. To ease comparison with the work of \citet{Henley}, we adopt the same values of $\Theta$ as done by these authors.  
\begin{table}
\caption{Parameters of the test models for circular binary systems with an orbital separation of 100\,R$_{\odot}$. All systems feature the same primary star, but differ by the properties of the secondary.}
\begin{tabular}{l c c c c}
\hline
Parameter & Primary & \multicolumn{3}{c}{Secondary} \\
\cline{3-5}
          &         & I & II & III \\
\hline
$\dot{M}$ ($10^{-6}$\,M$_{\odot}$\,yr$^{-1}$) & 1.0 & 0.98 & 0.3 & 0.1 \\
$v_{\infty}$ (km\,s$^{-1}$) & 2000 & 2000 & 2000 & 2000 \\
$R_*$ (R$_{\odot}$) & 20 & 20 & 10 & 10 \\
$M_*$ (M$_{\odot}$) & 30 & 30 & 30 & 30 \\
$\eta$            & -  & 0.98 & 0.3 & 0.1 \\
$\chi$            & $\geq 5.6$ & 5.7 & 13.1 & 26.8 \\
$\theta_{\infty}$($^{\circ}$) &  & 90.4 & 111.5 & 128.8 \\
\hline
\end{tabular}
\label{table1}
\end{table}

\begin{figure*}[htb]
\begin{center}
\includegraphics*[width=0.8\textwidth,angle=0]{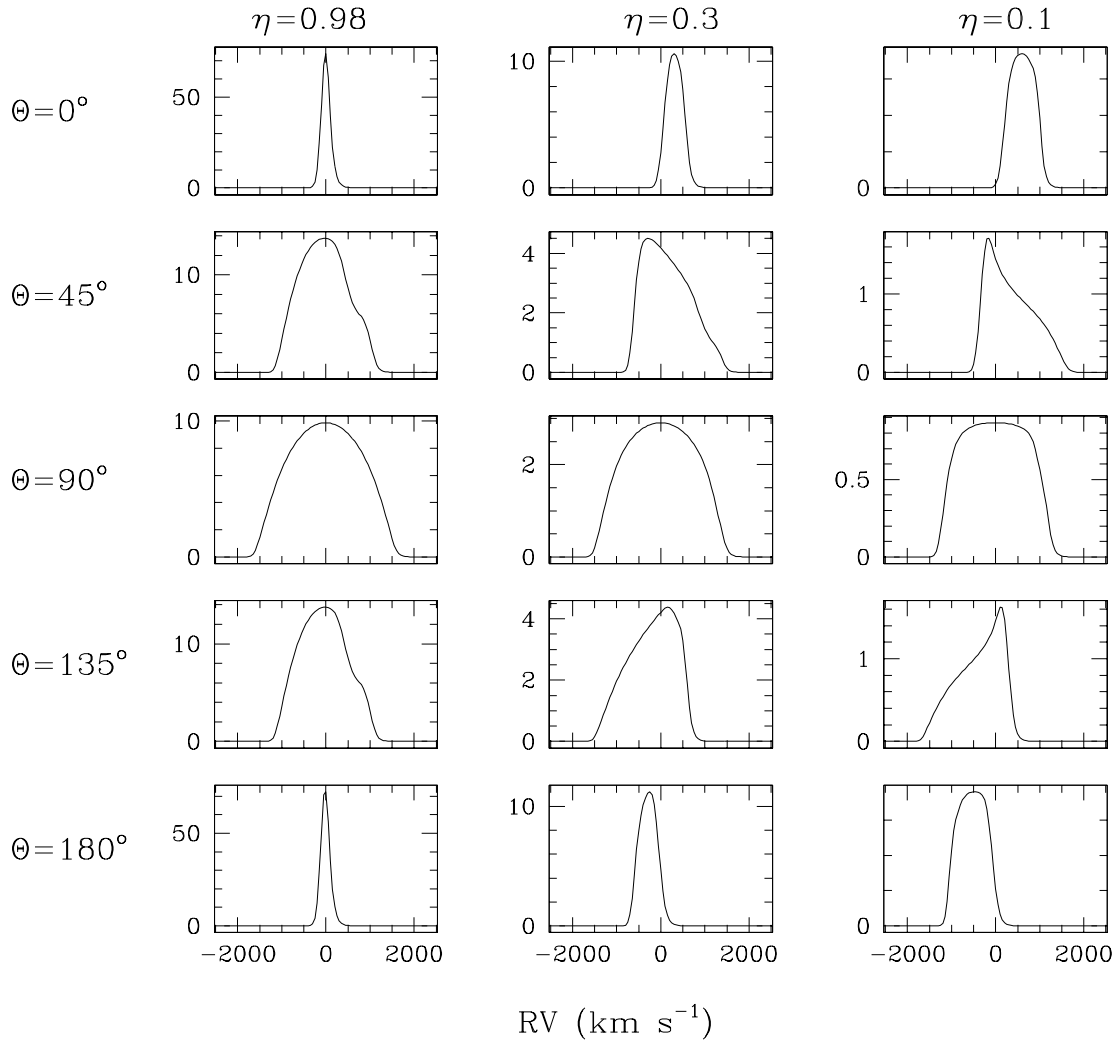}
\end{center}
\caption{Individual line profiles as a function of viewing angle $\Theta$ for the three colliding wind systems listed in Table\,\ref{table1}. The angle $\Theta$ is related to the orbital inclination and the true anomaly via equation\,\ref{eq:Theta}.\label{figure4}}
\end{figure*}
\begin{figure*}[htb]
\begin{center}
\includegraphics*[width=0.8\textwidth,angle=0]{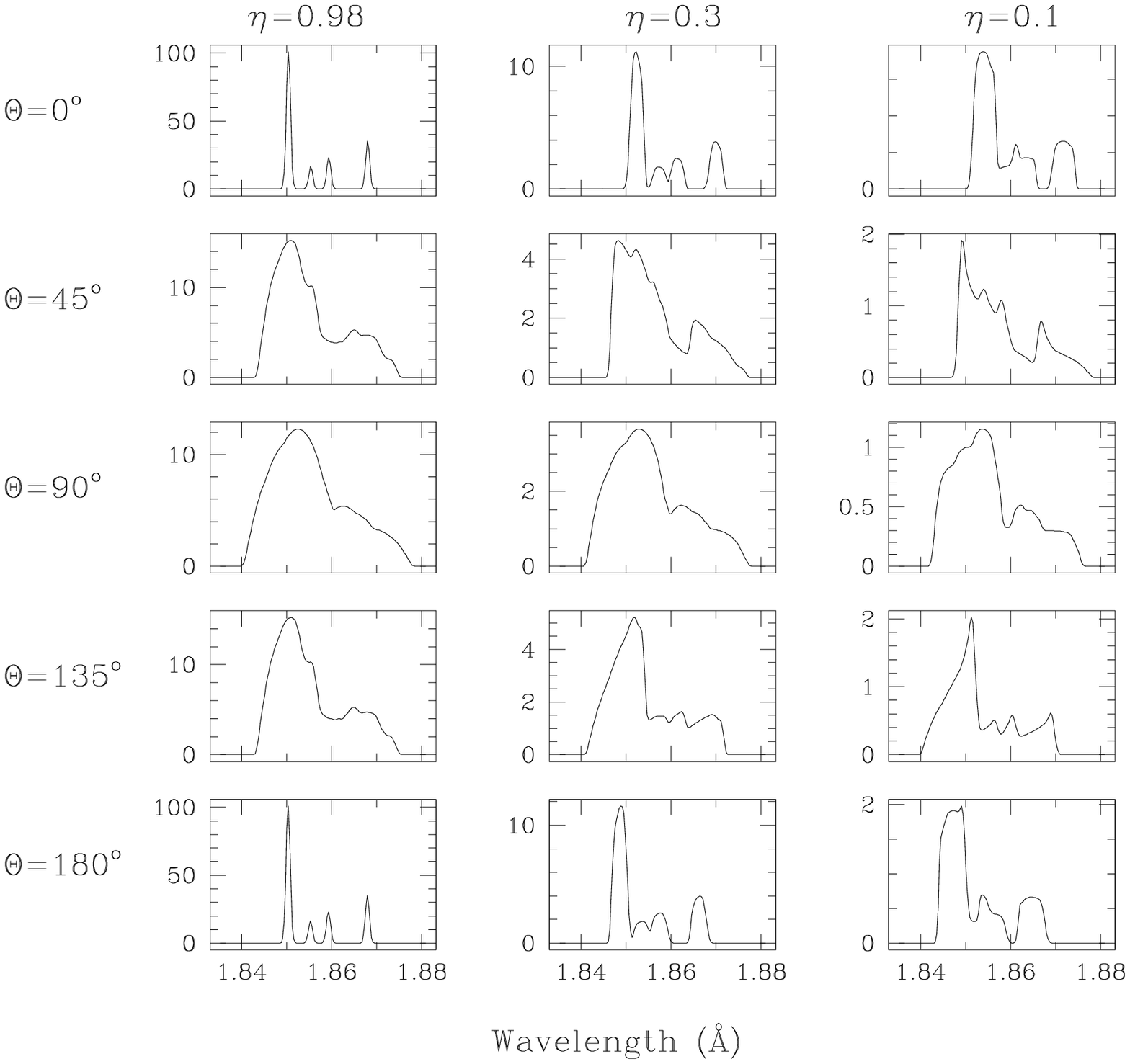}
\end{center}
\caption{The Fe {\sc xxv} line complex as a function of viewing angle $\Theta$ for the three colliding wind systems listed in Table\,\ref{table1}.\label{figure5}} 
\end{figure*}
Because of numerical singularities our code cannot deal with the case where $\eta$ is strictly equal to one. Case I thus rather considers $\eta = 0.98$. Figure\,\ref{figure4} illustrates the profiles of individual lines as a function of viewing angle for the three cases described in Table\,\ref{table1}. Figure\,\ref{figure5} illustrates the expected morphologies of the full Fe {\sc xxv} complex.

We start by considering Case I, which corresponds to a situation where the wind interaction zone is almost planar and located midway between the stars. Therefore, for conjunction phases ($\Theta = 0^{\circ}$ for the conjunction with the primary in front and $\Theta = 180^{\circ}$ for the opposite situation), the velocity vectors of the material flowing out of the wind interaction zone are essentially perpendicular to the line of sight. This results in very narrow profiles, centered on zero velocity. For these configurations, one hence expects to resolve the individual components of the Fe {\sc xxv} complex (see Fig.\,\ref{figure5}). We note that our results slightly underestimate the line width at these viewing angles. In a real system, the velocity vectors in the wind interaction zone are not simply tangent to the contact discontinuity, but their direction spans a range of values around the tangent. The hydrodynamic simulations of \citet{Henley} indeed yield somewhat wider lines for $\eta = 1$ and $\Theta = 0$ or $180^{\circ}$. For other viewing angles, Case I shows very large variations of the width of the line. The width is maximum for $\Theta = 90^{\circ}$, i.e.\ when the line of sight crosses the plane of the wind interaction region. For this orientation, the full-width of the line could reach $2 \times v_{\infty}$  if all cells of the wind interaction zone were emitting at the same level. The fact that the line width does not reach this value reflects the variations of the plasma temperature and hence emissivity along the shock: the further we move away from the binary axis, the more oblique the shock with respect to the inflowing wind and thus the lower the post-shock temperature and hence the lower the contribution to the line emission. We further see that the profile remains roughly symmetric and centered on zero velocity. Unlike the lines simulated by \cite{Henley}, the profiles in Fig.\,\ref{figure4} do not display strong asymmetries between the blue and red side. This is because the impact of photoelectric absorption by the cool unshocked winds is very small at the energies of the Fe {\sc xxv} lines considered here. Photoelectric absorption mainly affects the lines at longer wavelengths that are modelled by \citet{Henley}. Its net effect is to skew these lines towards the blue. In our simulations, the only slight asymmetries between the blue and red wings are seen for $\Theta = 45$ and $135^{\circ}$, where we note a slight depression in the red wing which stems from occultation of receding material by the star that is in front.

Another difference with the results of \citet{Henley} concerns the fact that, in their simulations, the unabsorbed profiles of the lighter elements display a double-peaked morphology for viewing angles away from conjunction. The reason for this behaviour is that the emissivities of the Ly$\alpha$ lines of lighter elements peak at some distance away from the binary axis, leading to an emission region that has essentially the shape of a cone with its apex truncated \citep[this is similar to the situation of the optical line profiles modelled by][]{Luehrs}. Conversely, the emissivity of Fe {\sc xxv} reaches its maximum near the apex and progressively decreases along the cone. This situation leads to profiles that exhibit a single maximum peak near the line-of-sight velocity of the apex.  

We finally note that the strong line broadening for viewing angles away from conjunction leads to severe blending of the various Fe {\sc xxv} components. The resulting profiles are thus rather complex (see Fig.\,\ref{figure5}). In practice, this result highlights the need to implement a least-squares deconvolution technique \citep[LSD,][]{LSD} to recover the profiles of individual lines from observed data. In the remainder of this paper, we will mainly focus on the individual line profiles, assuming that they have been extracted via LSD.\\  

A different situation prevails for $\eta = 0.3$ and $\eta = 0.1$ (Cases II and III, respectively). At conjunction phases, we still observe a relatively narrow line, though it is now much wider than in Case I. We also note that the profile, although narrow, is now markedly skewed towards the red for $\Theta = 0^{\circ}$ and towards the blue for $\Theta = 180^{\circ}$. This is of course because the wind interaction zone is now wrapped around the star with the weaker wind (the secondary) and the post-shock material is thus moving away from the observer at $\Theta = 0^{\circ}$ and moving towards the observer at $\Theta = 180^{\circ}$. Again, the maximum width (for the five values of $\Theta$ considered here) is observed at quadrature ($\Theta = 90^{\circ}$), where our calculations predict a symmetric profile centered on the rest wavelength of the lines. At intermediate phases, highly asymmetric line profiles with a marked peak near zero velocity are predicted, especially for Case III. This peak results from the fact that at these values of $\Theta$ and for the specific value of $\theta_{\infty}$, one of the arms of the interaction region is almost perpendicular to the observer's sightline.\\ 

\begin{figure*}[htb] 
\begin{center}
\includegraphics*[width=0.3\textwidth,angle=0]{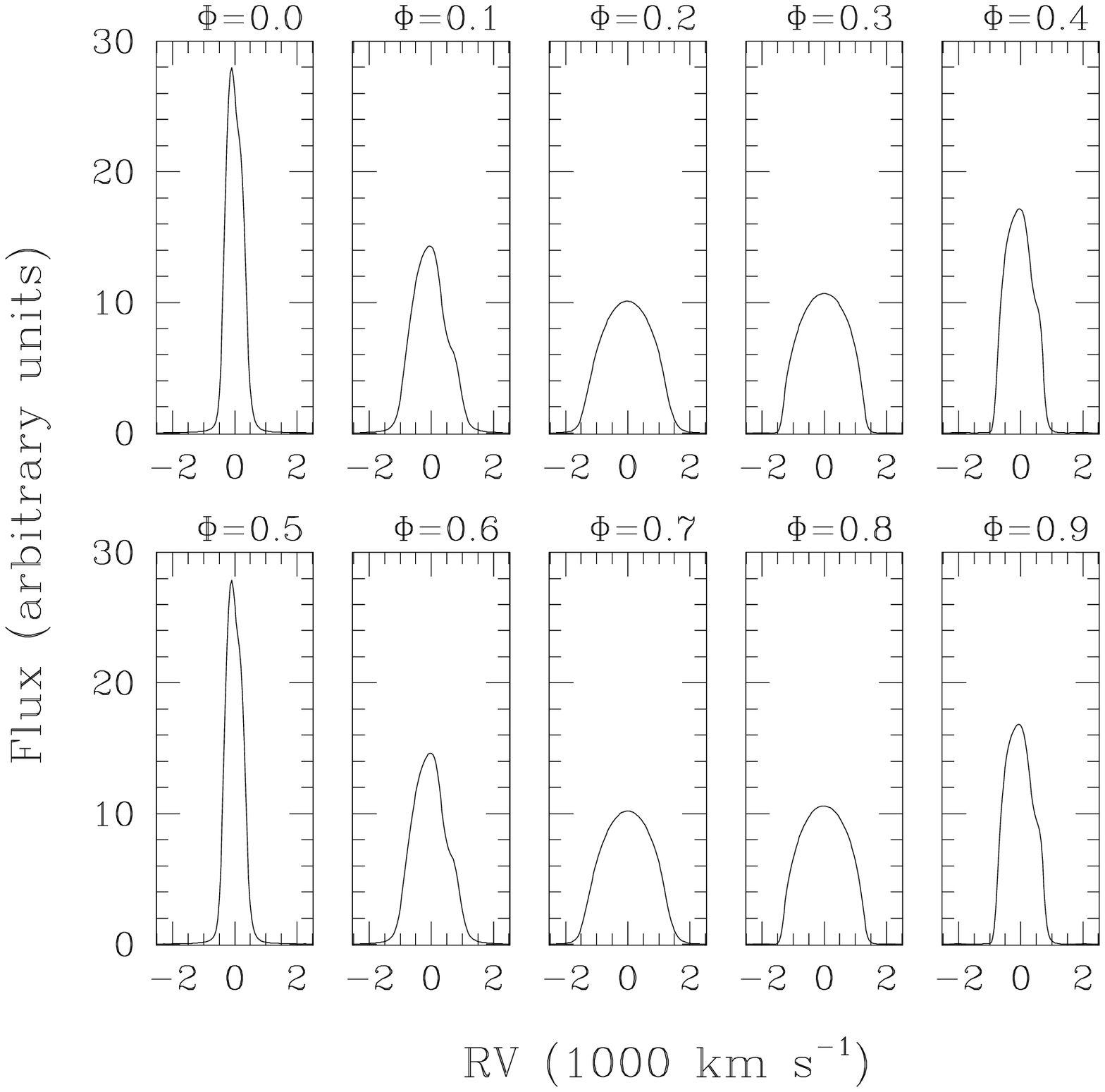}
\includegraphics*[width=0.3\textwidth,angle=0]{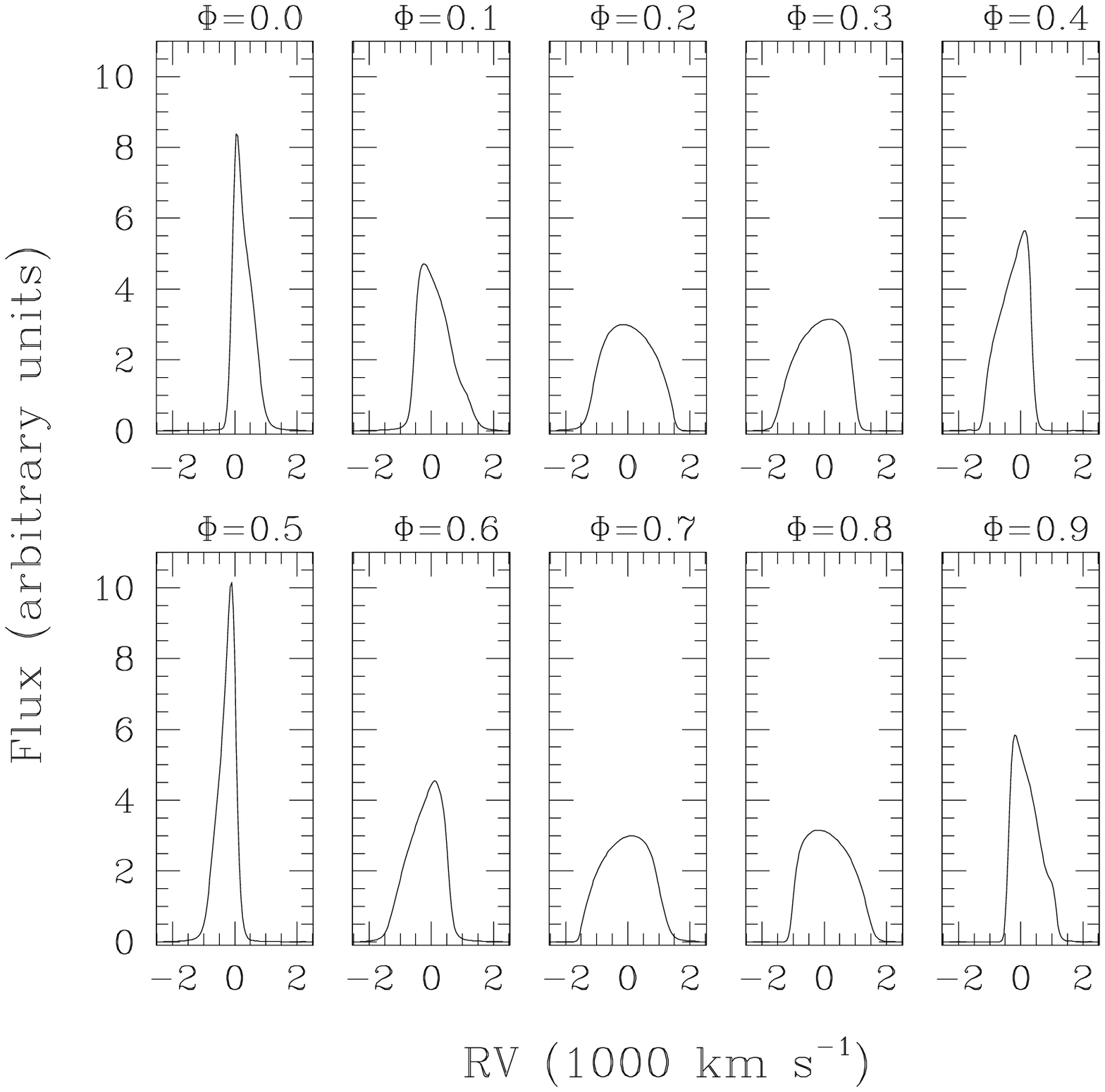}
\includegraphics*[width=0.3\textwidth,angle=0]{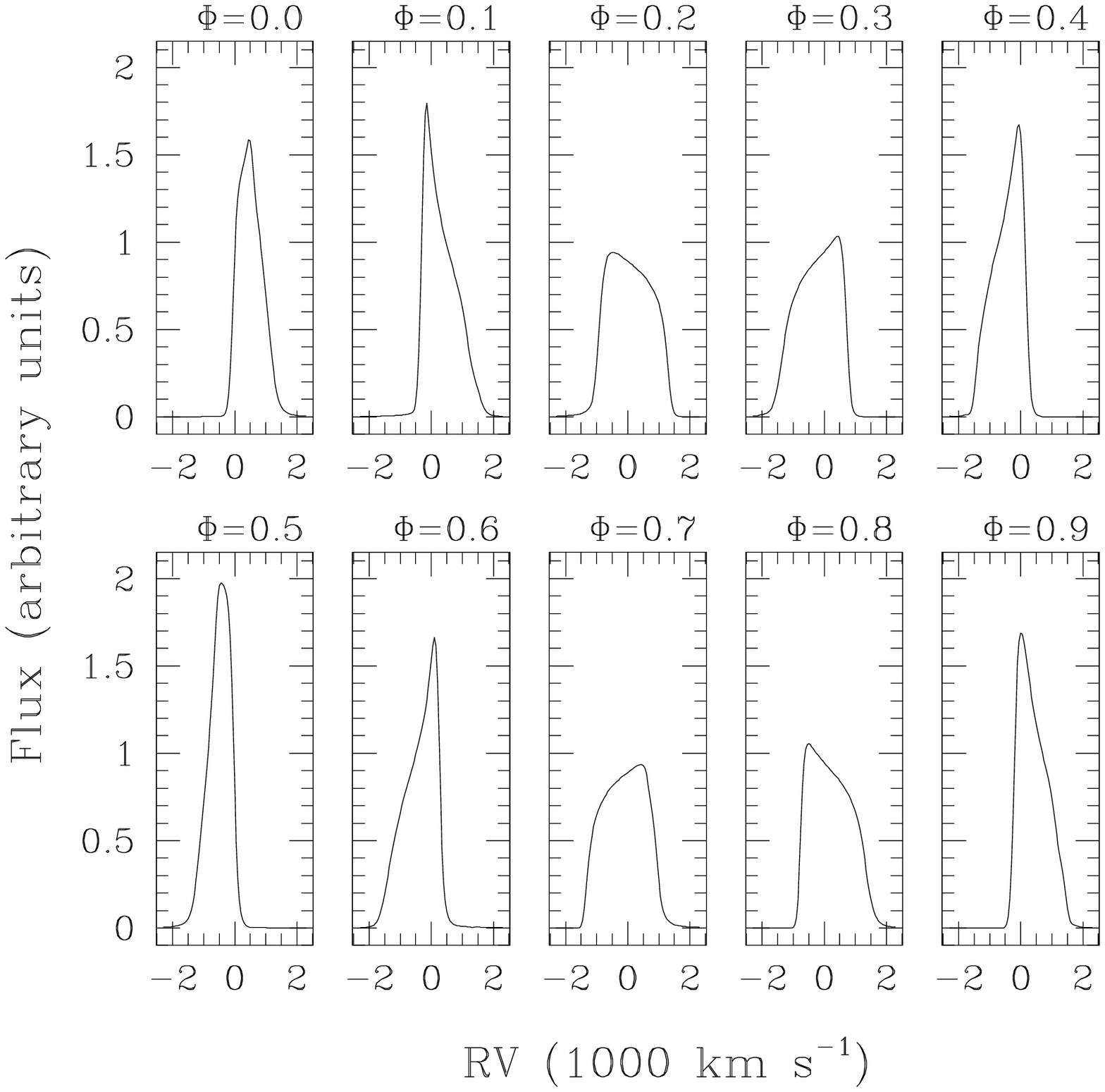}
\end{center}
\begin{center}
\includegraphics*[width=0.3\textwidth,angle=0]{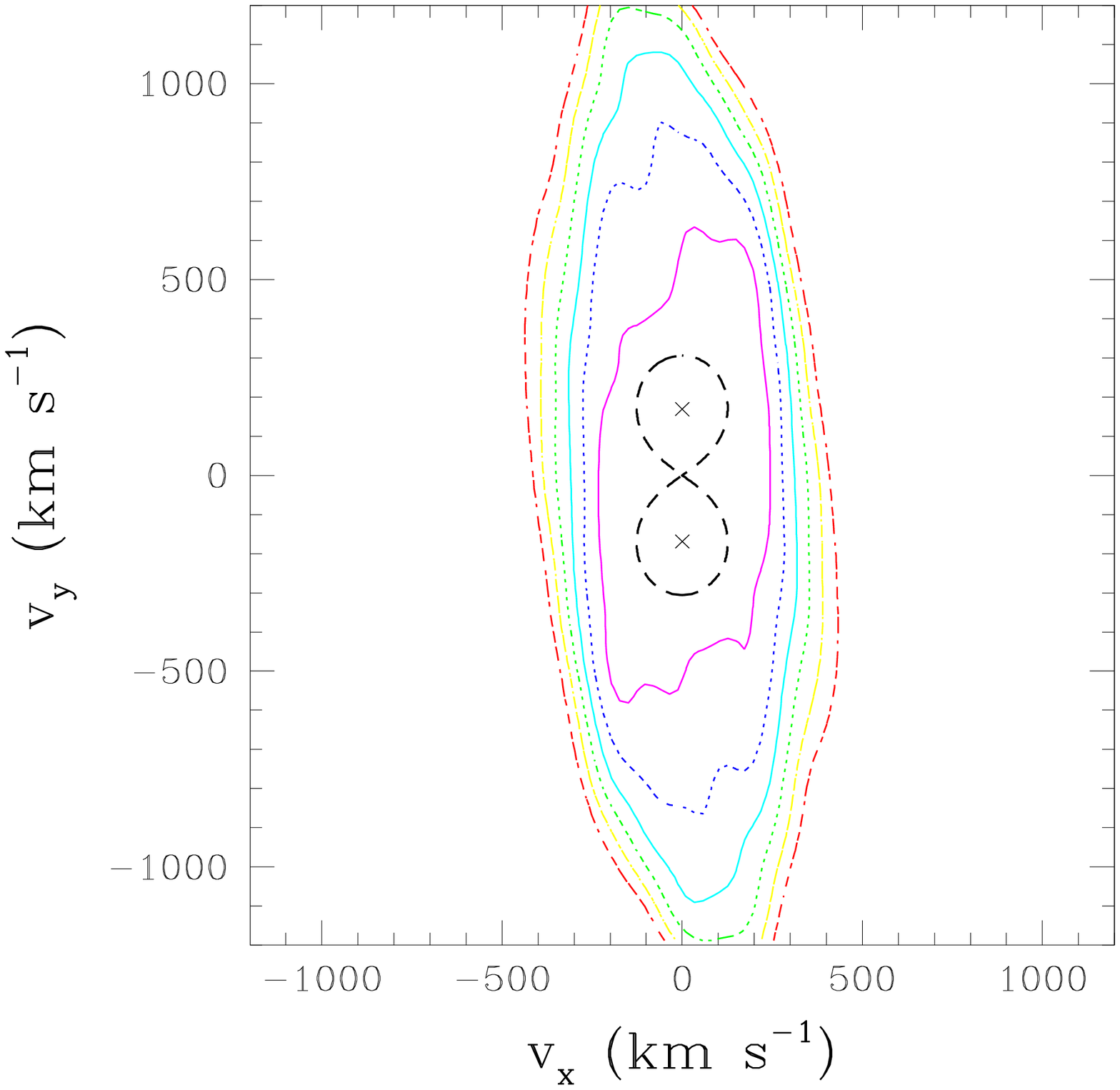}
\includegraphics*[width=0.3\textwidth,angle=0]{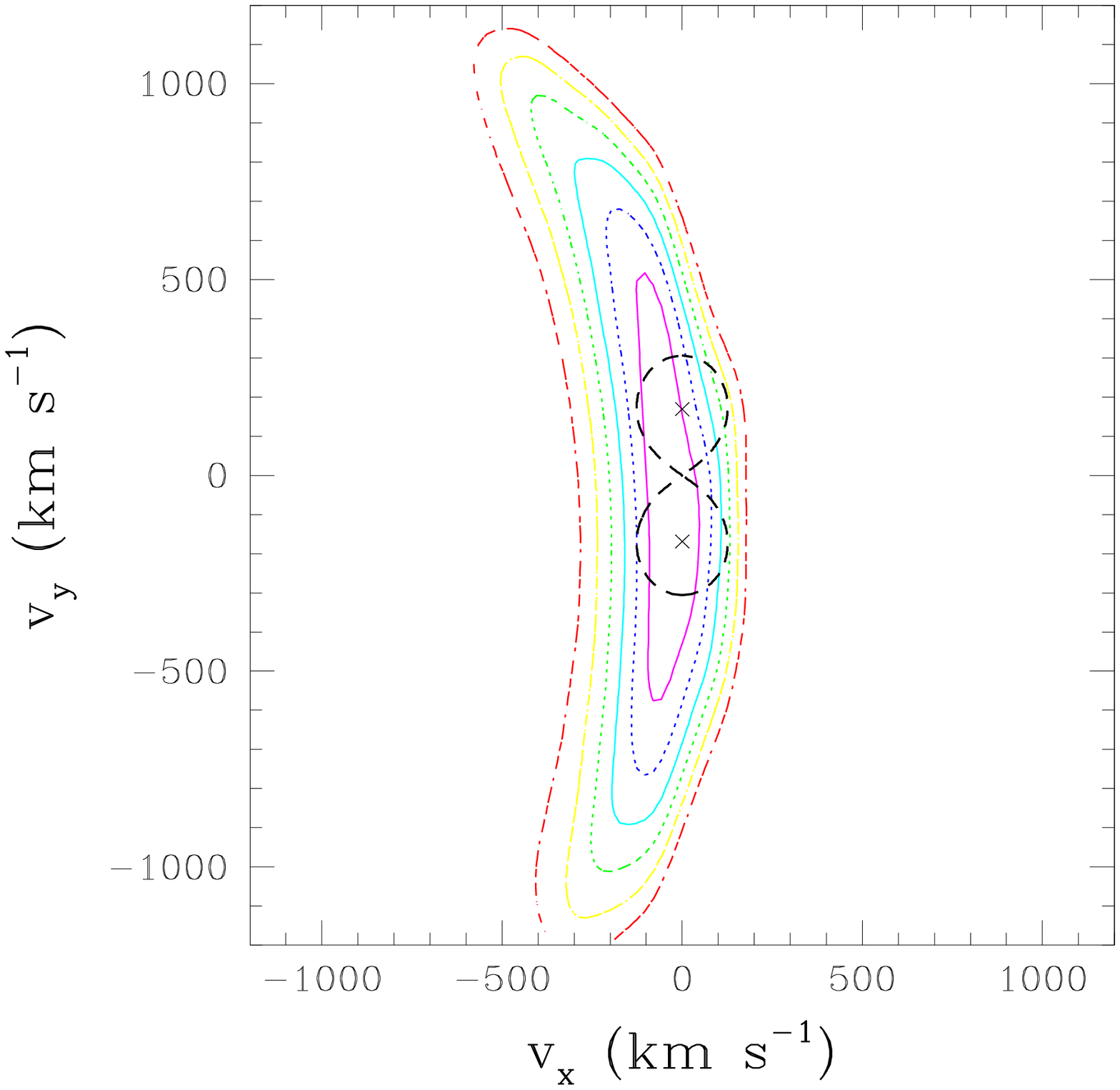}
\includegraphics*[width=0.3\textwidth,angle=0]{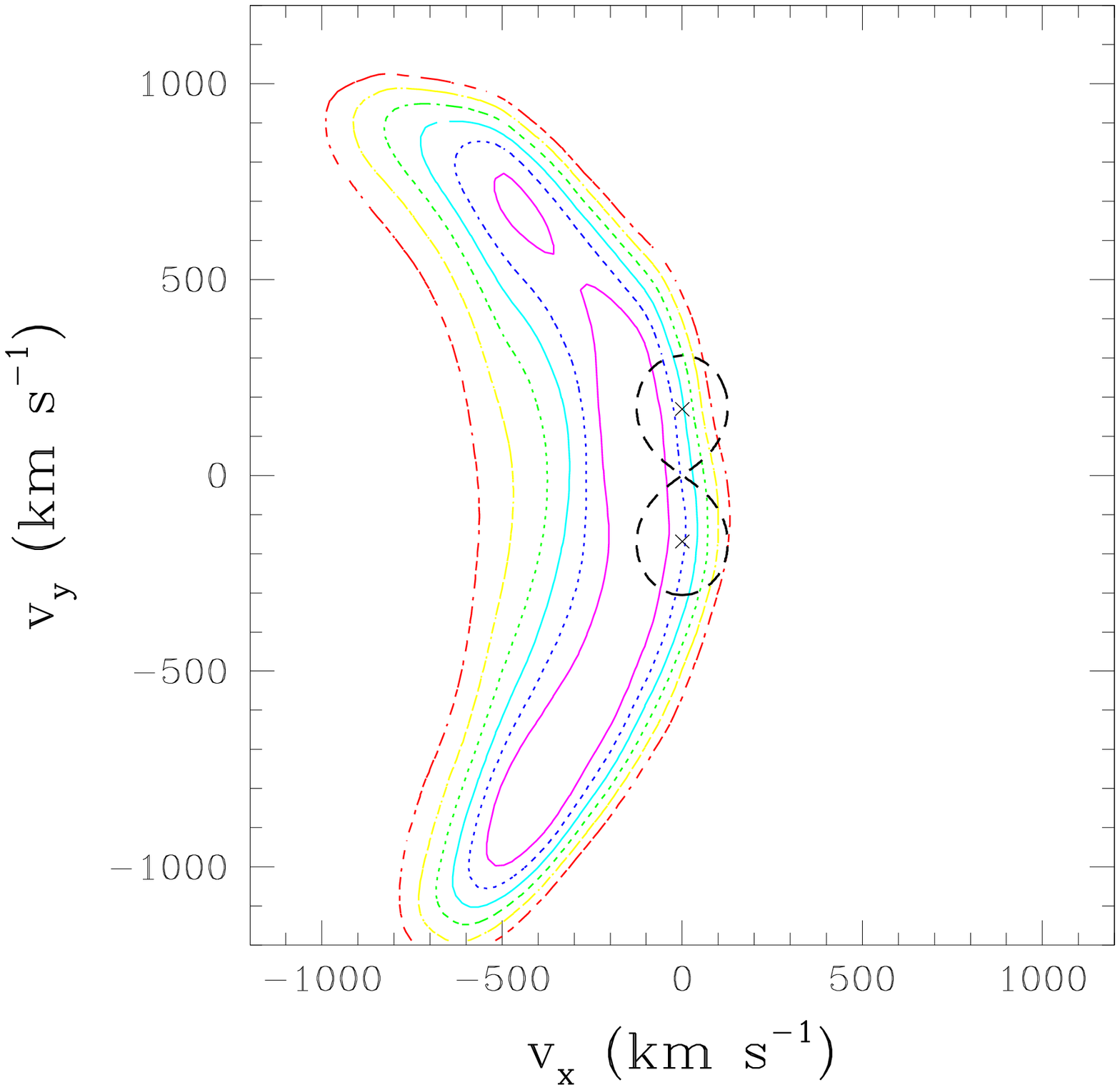}
\end{center}
\caption{Top row: phase-resolved line profiles for the resonance line of the Fe {\sc xxv} complex for the three systems listed in Table\,\ref{table1} seen under $i = 75^{\circ}$ (from left to right: Case I, Case II and Case III). Bottom row: corresponding Doppler maps computed from 20 simulated line profiles for each system. The levels of the various contours correspond to 90, 80, 70, 60, 50 and 40\% of the peak intensity in the map. The Roche lobes in velocity space are shown by the dashed contours, whilst the crosses indicate the orbital velocity amplitudes of the primary and secondary stars. \label{figure6} }
\end{figure*}

As a next step, we have simulated phase-resolved synthetic line profiles and spectra for the three cases above, but now assuming binary systems seen under an orbital inclination of $i = 75^{\circ}$. At this stage, we also account for the orbital motion by requesting the wind interaction zone to rotate with the stars on their orbit. The resulting profiles of individual lines at ten distinct orbital phases are shown in the top row of Fig.\,\ref{figure6}. The variations of the line profiles are clearly seen. In the following subsection, we analyse these profile variations by means of the so-called Doppler tomography technique.

\subsection{Doppler tomography \label{tomography}}
Doppler tomography is an elegant and powerful technique to study the dynamics of gas flows in spectroscopic binaries with circular orbits \citep{Horne,kaitchuk}. This method is nowadays widely applied to optical spectra, but to the best of our knowledge, it has so-far never been used in the X-ray domain.

Doppler tomography translates phase-locked emission line profile variations into a map of the line formation region in velocity space. To do so, one adopts a reference frame centered on the centre of mass of the binary with the $x$-axis pointing from the secondary to the primary and the positive $y$-axis pointing along the direction of the primary's orbital motion. The fundamental assumption of the Doppler tomography technique is that the phase dependence of the radial velocity $v(\Phi)$ of any cell of emitting gas that is stationary in the rotating frame of reference of the binary can be described by the simple relation:
\begin{equation}
v(\Phi) = -v_x\,\cos{(2\,\pi\,\Phi)} + v_y\,\sin{(2\,\pi\,\Phi)}
\label{eq:4}
\end{equation}
Here $\Phi$ stands for the orbital phase, with $\Phi = 0$ at conjunction with the primary star in front. The pair $(v_x, v_y)$ yields the velocity coordinates of the gas cell projected along the $x$ and $y$ axes: $v_x = V_x\,\sin{i}$ and $v_y = V_y\,\sin{i}$, where $V_x$ and $V_y$ are the actual velocities in the orbital plane of the binary and $i$ is the orbital inclination. 

The Doppler map $DM(v_x,v_y)$ yields a measure of the flux that is carried across the line profile for each particular $(v_x,v_y)$ pair. If we adopt the back-projection method, the Doppler map can be expressed as 
\begin{equation}
DM(v_x,v_y) = \frac{\int D(v(v_x,v_y,\Phi),\Phi)\,W(\Phi)\,d\Phi}{\int W(\Phi)\,d\Phi}
\label{eq:5}
\end{equation}
where $D(v,\Phi)$ is the observed flux density at radial velocity $v(v_x,v_y,\Phi)$ (given by equation\,\ref{eq:4}) at phase $\Phi$. $W(\Phi)$ is the weight assigned to the observation at phase $\Phi$ \citep[see e.g.][for a detailed discussion of the method]{Horne}. 
  
The implementation of the Doppler tomography that we use here is adapted from the one used by \citet{WR20a}. The method is based on a Fourier filtered back-projection algorithm \citep{Horne}. The Point Spread Function (PSF) of a pure back-projection technique in the $(v_x,v_y)$ plane has a Gaussian core with extended wings having a $1/\sqrt{v_x^2 + v_y^2}$ profile \citep{Horne}. To sharpen the PSF we apply a Fourier filter to the spectra prior to the back projection. For this purpose, we first compute the Fourier transform of the trailed spectrogram and we then multiply the result by a filter $\frac{\omega}{\omega_N}\,\exp{\left[-\frac{(\omega/\omega_C)^2}{2}\right]}$ to suppress the $1/\sqrt{v_x^2 + v_y^2}$ tail and to prevent the amplification of high-frequency noise in the spectra, where $\omega_N$ is the Nyquist frequency of the spectra, $\omega_C$ is set by the spectral resolution \citep[for further details see e.g.][]{Horne}. Finally, an inverse Fourier transformation is performed to recover the filtered spectra that are then back-projected. 

Back-projection produces stripes across the Doppler map at angles corresponding to the binary phases sampled by the data. Therefore, the back-projection method can lead to the so-called `radial-spoke artefact' if the data do not provide a uniform coverage of the orbital cycle. To avoid this problem, we have sampled our synthetic spectra at 20 equally-spaced orbital phases. 

The results of applying our method to the three sets of simulations are shown in Fig.\,\ref{figure6}. It becomes clear from these examples that the aspect of the Doppler map strongly depends on the shock opening angle and hence the value of the wind momentum ratio $\eta$. Doppler tomography of the Fe {\sc xxv} line thus provides a sensitive diagnostics of the relative strengths of the winds in colliding wind binaries with circular orbits. 

\begin{table}
\caption{Parameters of the test models for eccentric binary systems.}
\begin{tabular}{l c c c c}
\hline
Parameter & \multicolumn{2}{c}{IV} & \multicolumn{2}{c}{V}\\
\cline{2-3}\cline{4-5}
          & Prim. & Seco. & Prim. & Seco. \\
\hline
$\dot{M}$ ($10^{-6}$\,M$_{\odot}$\,yr$^{-1}$) & 12.0 &  5.0 & 57.0 & 1.8 \\
$v_{\infty}$ (km\,s$^{-1}$)                  & 2400 & 2500 & 2860 & 3200 \\
$R_*$ (R$_{\odot}$) & 16 & 19 & 13 & 12 \\
$M_*$ (M$_{\odot}$) & 44 & 50 & 16 & 41 \\
$i$($^{\circ}$) & \multicolumn{2}{c}{62} & \multicolumn{2}{c}{55} \\
$e$ & \multicolumn{2}{c}{0.71} & \multicolumn{2}{c}{0.896} \\ 
$\omega$($^{\circ}$) & 12 & 192 & 224.6 & 44.6 \\
$a$ (R$_{\odot}$)    & \multicolumn{2}{c}{1727}& \multicolumn{2}{c}{3300} \\
$\eta$            & \multicolumn{2}{c}{0.43} & \multicolumn{2}{c}{0.035} \\
$\chi$            & $\geq 5.8$ & $\geq 10.8$ & $\geq 2.4$ & $\geq 21.9$ \\
$\theta_{\infty}$($^{\circ}$) & \multicolumn{2}{c}{105.1} & \multicolumn{2}{c}{142.3} \\ 
\hline
\end{tabular}
\label{table2}
\end{table}

\subsection{Systems with eccentric orbits}
In this section we use our code to simulate line profiles for a sample of eccentric colliding wind binary systems. In addition to the wind parameters, the orbital inclination and major axis, such systems are characterized by two new parameters: the eccentricity $e$ and the longitude of periastron $\omega$. 

\begin{figure*}[htb]
\begin{center}
\includegraphics*[width=0.3\textwidth,angle=0]{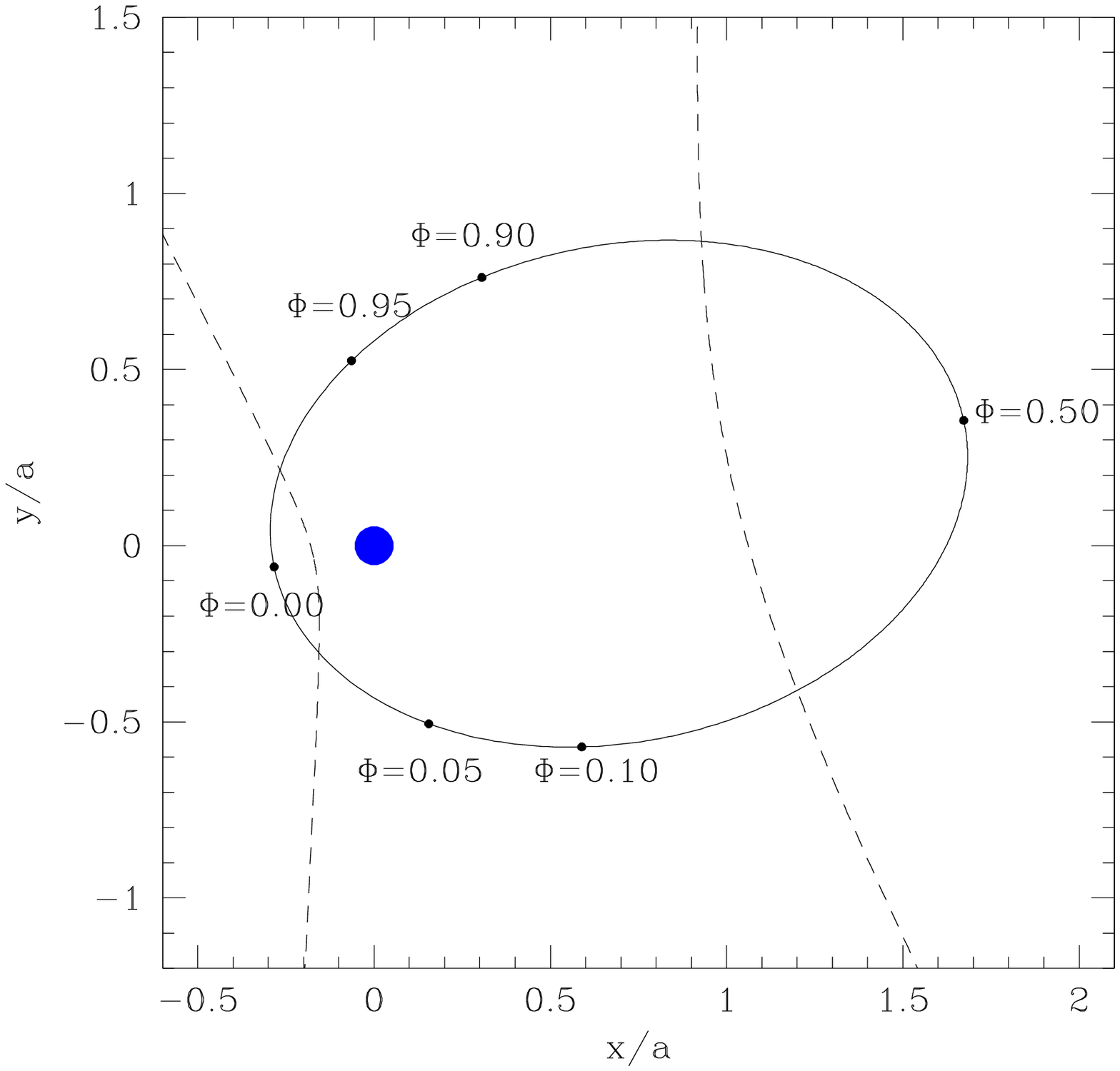}
\includegraphics*[width=0.3\textwidth,angle=0]{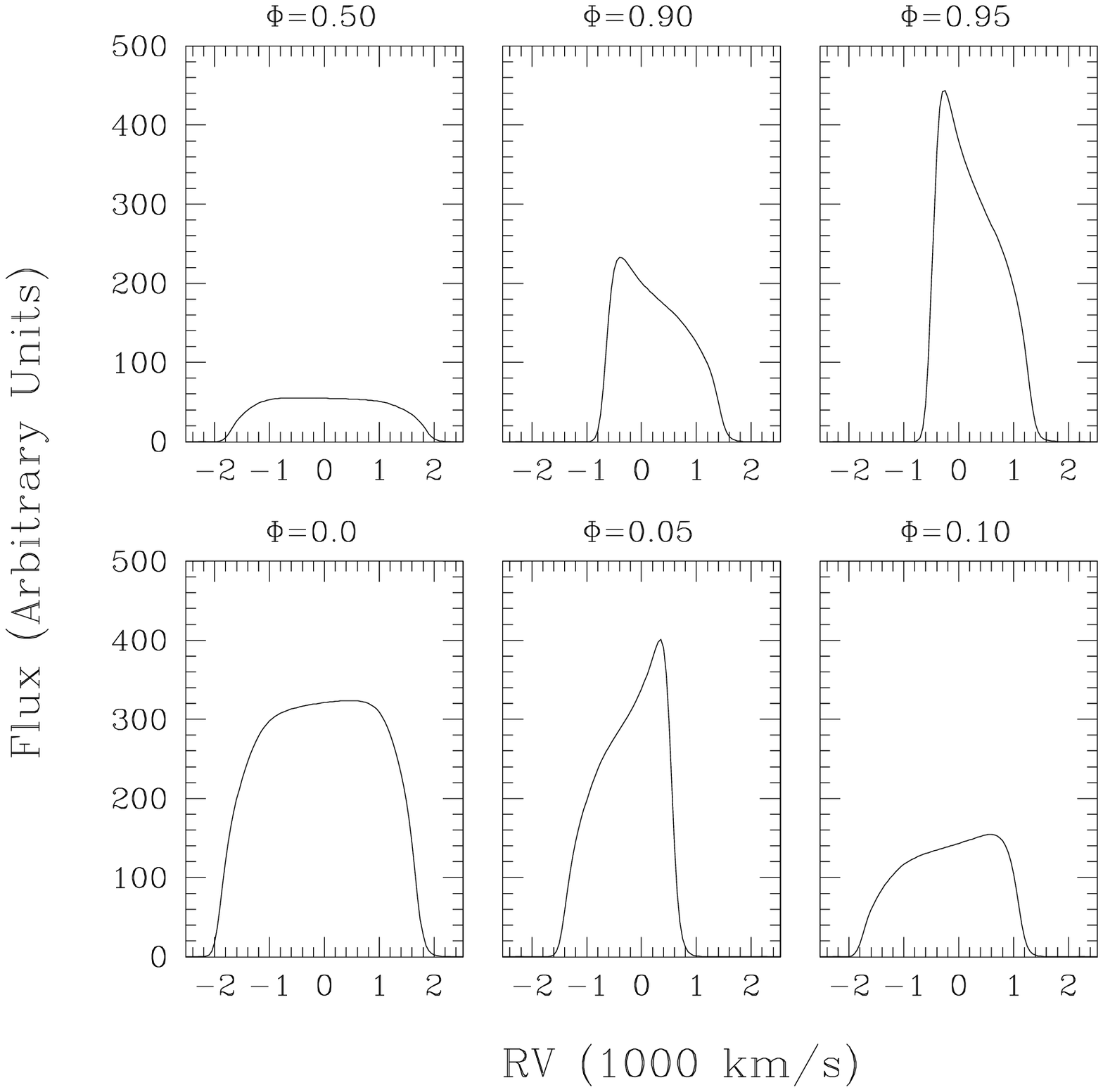}
\includegraphics*[width=0.3\textwidth,angle=0]{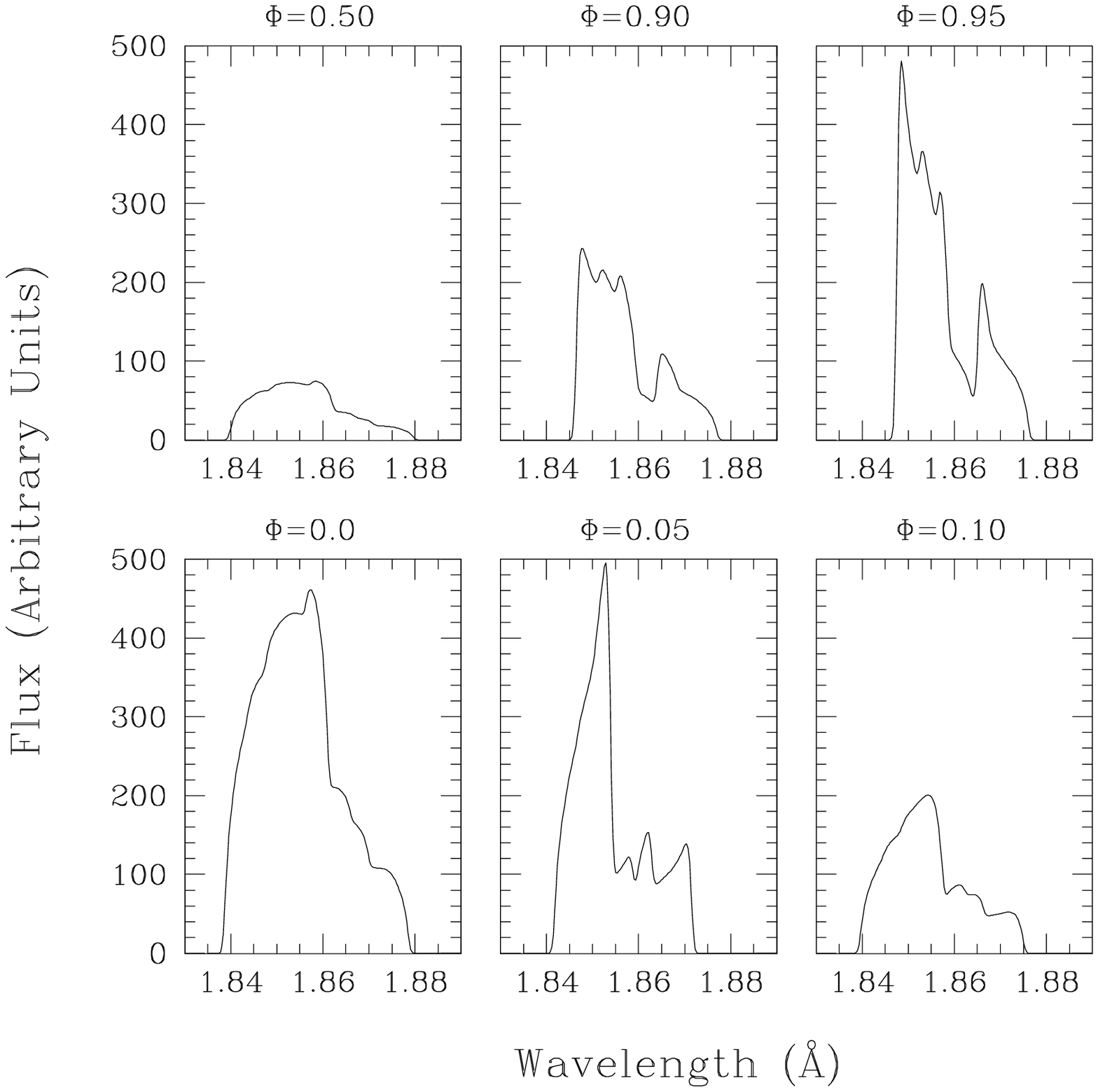}
\end{center}
\begin{center}
\includegraphics*[width=0.3\textwidth,angle=0]{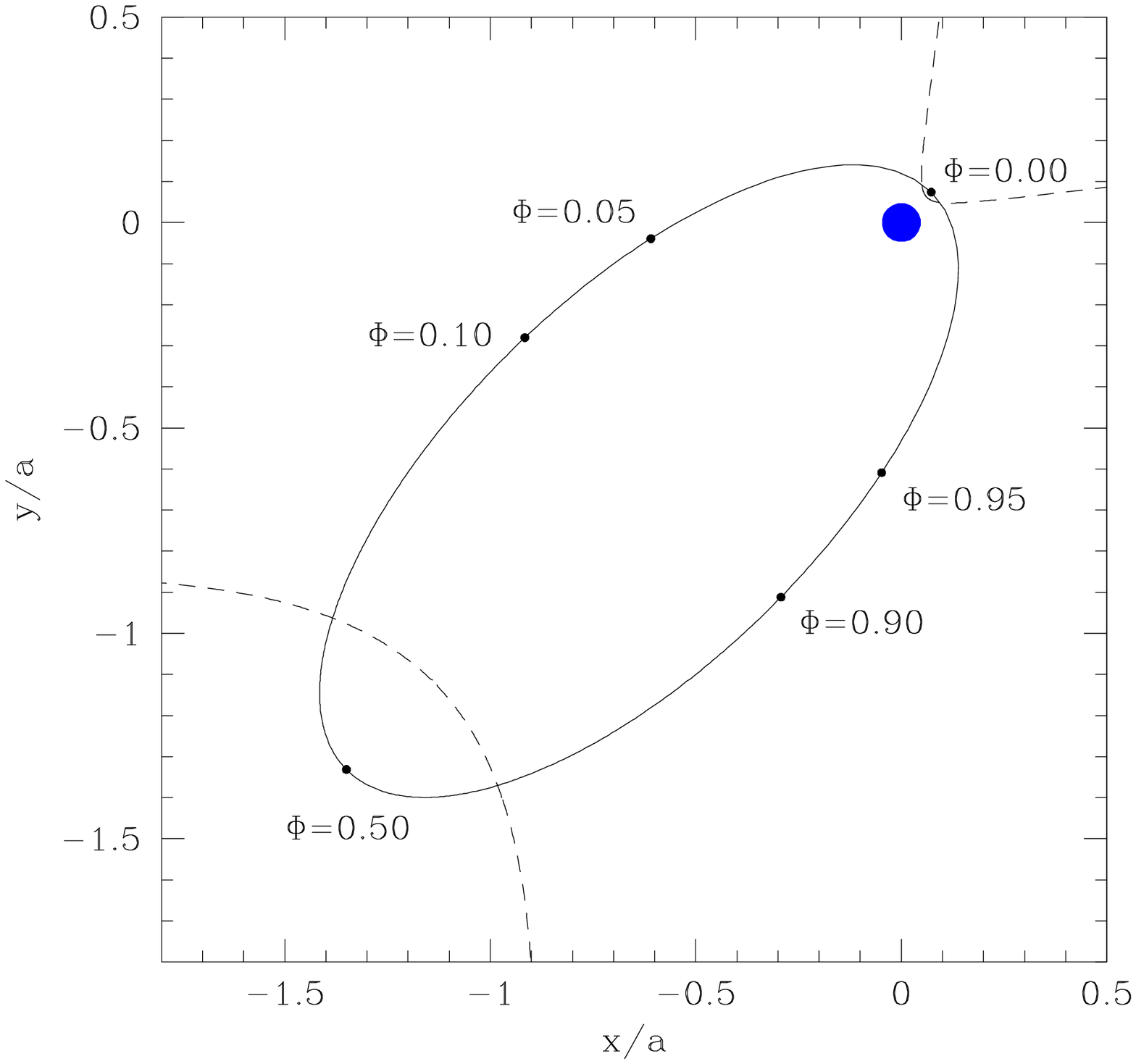}
\includegraphics*[width=0.3\textwidth,angle=0]{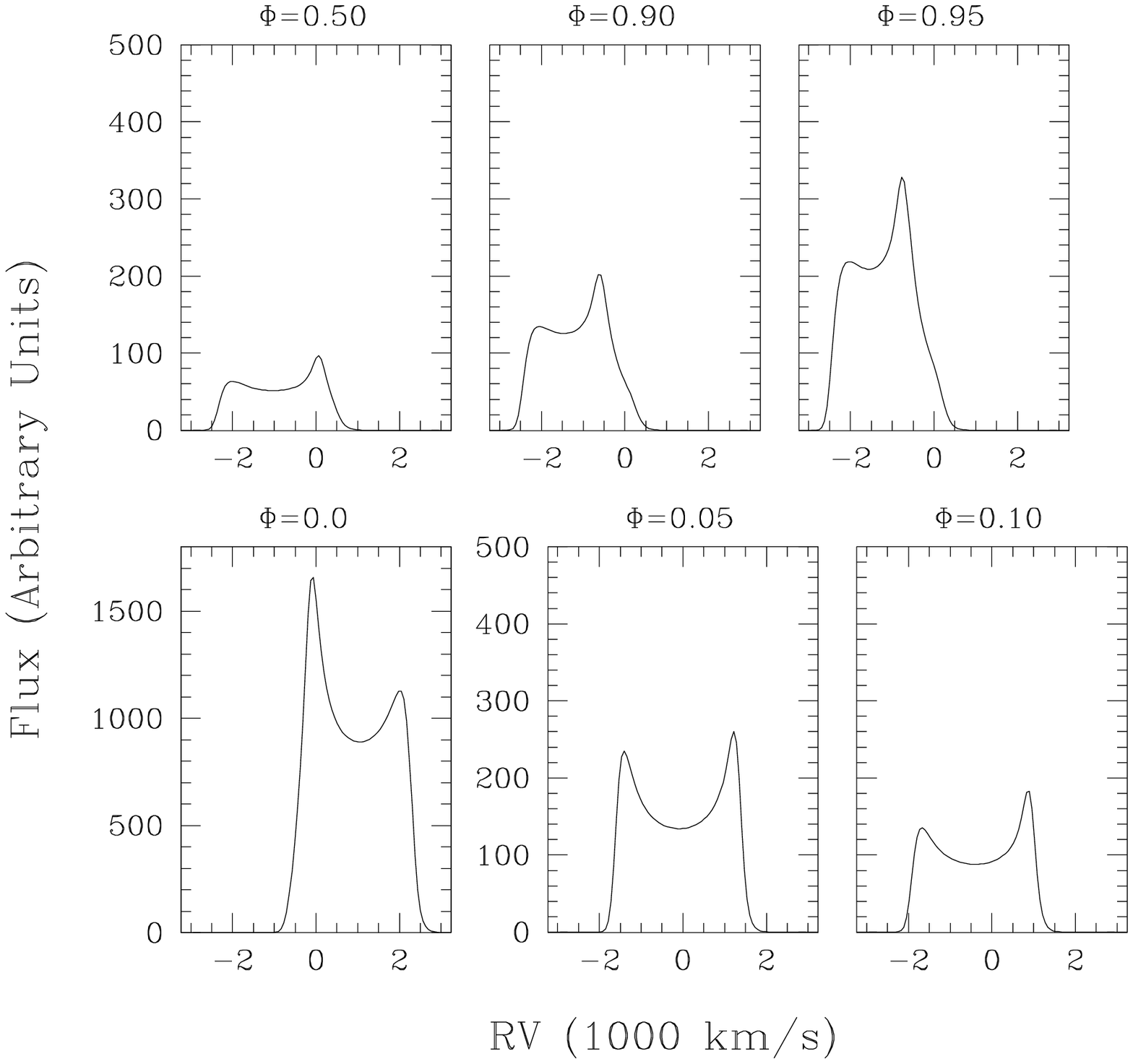}
\includegraphics*[width=0.3\textwidth,angle=0]{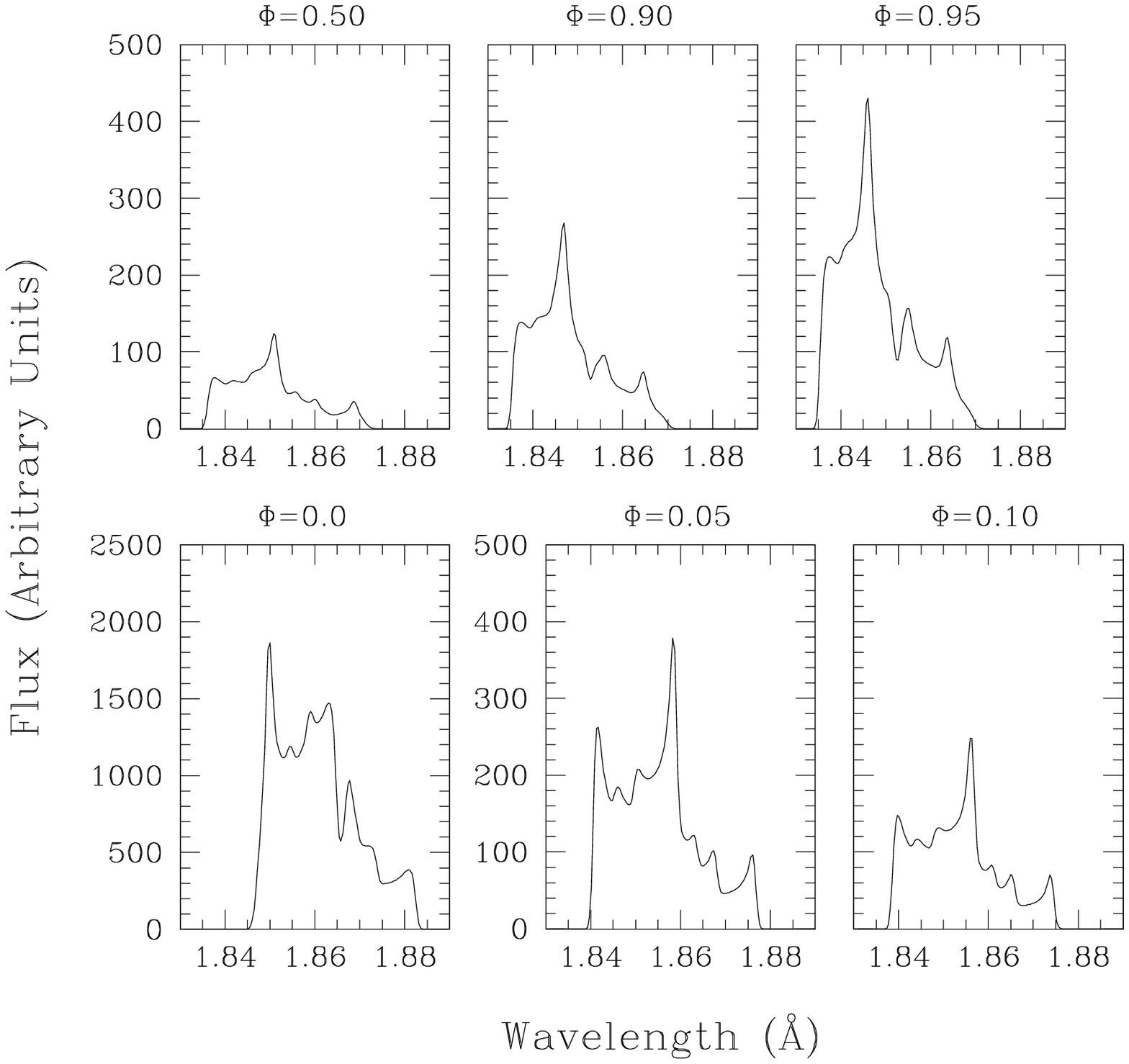}
\end{center}
\caption{Top left: sketch of the orbit of the elliptical system of Case IV. The star with the stronger wind is depicted by the big blue dot, whilst the small dots illustrate the position of the secondary at six orbital phases. The shape of the contact discontinuity computed according to the \citet{Canto} model is shown at periastron and apastron. The observer is looking up from the bottom at an inclination of $i = 62^{\circ}$. Top middle: synthetic individual profiles computed for Case IV and for the orbital phases shown on the left. Top right: synthetic Fe {\sc xxv} profiles for Case IV at the same orbital phases. Bottom: same as top, but for Case V. The orbital inclination is $55^{\circ}$ in this case. For clarity, all panels of the middle and right have the same vertical scale except for the periastron phase.\label{figure7}}
\end{figure*}

Table\,\ref{table2} lists the parameters of the systems that we consider here. They are directly inspired from Cyg\,OB2\,\#9 \citep[Case IV, see][]{Naze} and WR\,140 \citep[Case V, see][]{Pollock,Fahed}. 

For the eccentric systems, we predict strong variations of the integrated line flux with orbital phase. In fact, since we are dealing with simulations that assume adiabatic wind interaction zones, we actually recover the $1/d$ flux variation that is expected for such systems \citep{SBP}. 

Figure\,\ref{figure7} illustrates the results of our simulations for the two eccentric systems. For Case IV ($e = 0.71$, $\eta = 0.43$), the shock opening angle is relatively large, leading to rather broad lines at most orbital phases. The individual profiles have a single-peaked morphology. Their intensity, centroid and skewness all change with orbital phase. The full Fe {\sc xxv} complex displays a rather complicated morphology indicating that LSD might again be needed to disentangle the variations of the individual profiles from apparent variations due to the blending. 

For Case V ($e = 0.896$, $\eta = 0.035$), the primary wind overwhelms that of the secondary and, in our simple model, it is only the wide separation of the stars that allows for the shock region to remain detached from the secondary's surface. The adopted pre-shock velocities are very high, leading to post-shock temperatures near 139\,MK. This is above the temperature of maximum emissivity of the Fe {\sc xxv} complex. Hence, in our simulation, the maximum of the emission of the line occurs at positions slightly away from the shock apex, thereby leading to a double-peaked line morphology (see Fig.\,\ref{figure7}), as discussed in Sect.\,\ref{circular}. This renders the resulting profiles of the full Fe {\sc xxv} blend even more complex.

\begin{figure*}[thb]
\begin{center}
\includegraphics*[width=0.75\textwidth,angle=0]{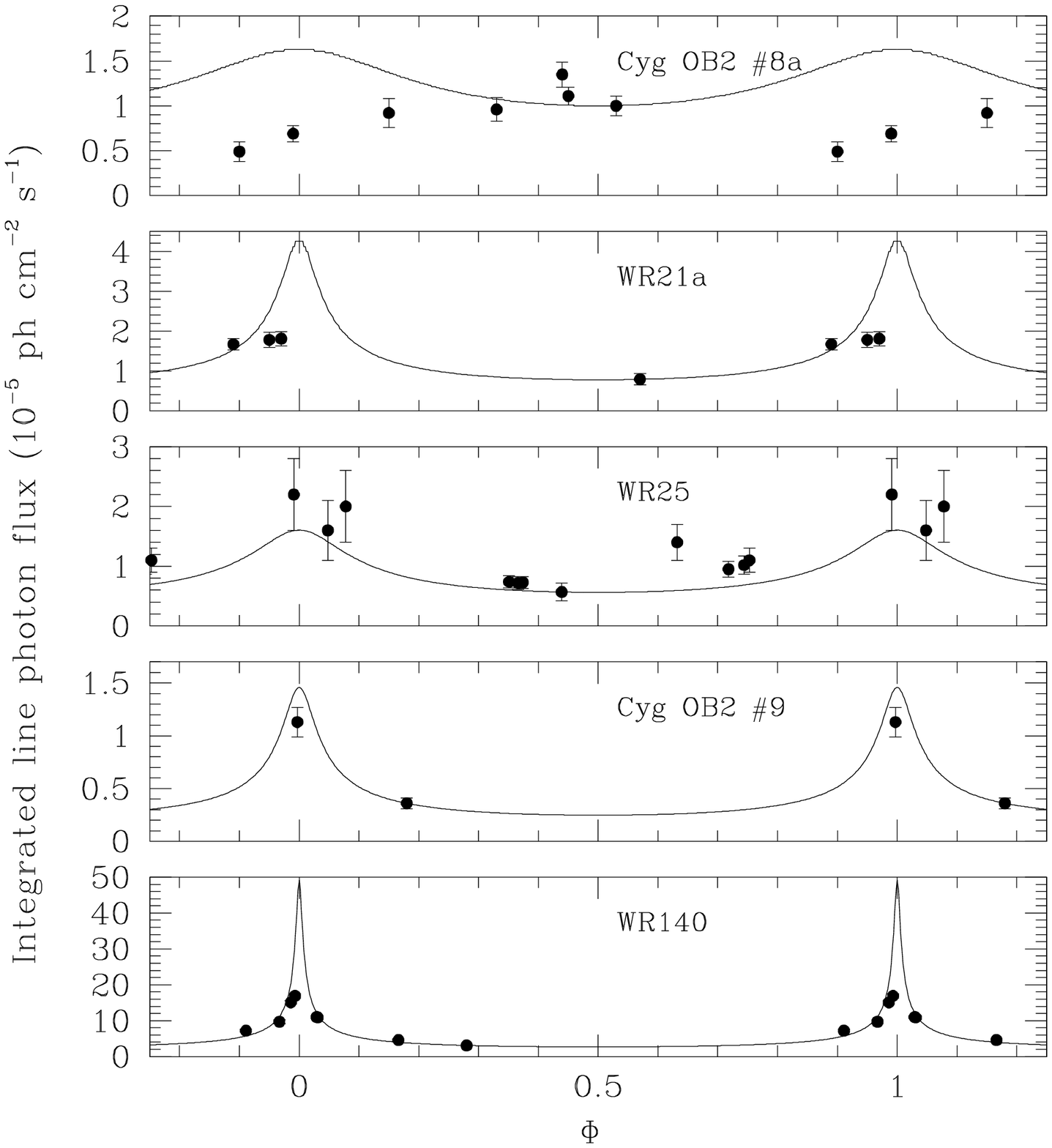}
\end{center}
\caption{Integrated photon fluxes of the Fe K line in five colliding wind binary systems in order of increasing orbital period from top to bottom. The symbols with the 1-$\sigma$ errorbars indicate the observed values whilst the solid lines indicate the $1/r$ trends expected for an adiabatic wind interaction zone scaled to the flux measured on the spectrum closest to apastron ($\Phi = 0.5$).\label{figure8}}
\end{figure*}

\subsection{Comparison with observations}
Current instruments in X-ray astrophysics lack the spectral resolution in the hard energy band needed to confront the profiles obtained in our simulations with real observations. Yet, we can compare the relative variations of the integrated line fluxes of several eccentric systems with the predictions of models for adiabatic wind interaction zones. For this purpose, we have considered five systems which exhibit a clear Fe {\sc xxv} feature in their broadband X-ray spectra. In increasing order of the orbital period, these are Cyg\,OB2 \#8a \citep[$P_{\rm orb} = 21.9$\,days, $e = 0.24$][]{DeBecker}, WR\,21a \citep[$P_{\rm orb} = 31.7$\,days, $e = 0.69$][]{Tramper}, WR\,25 \citep[$P_{\rm orb} = 208$\,days, $e = 0.48$][]{Gamen}, Cyg\,OB2 \#9 \citep[$P_{\rm orb} = 2.36$\,yrs, $e = 0.71$][]{Naze}, and WR\,140 \cite[$P_{\rm orb} = 7.9$\,yrs, $e = 0.89$][]{Fahed}. We have analysed {\it XMM-Newton} spectra for these systems. For Cyg\,OB2 \#8a, Cyg\,OB2 \#9 and WR\,21a, the description of the data is given by \citet{Cazorla}, \citet{Naze}, and \citet{Gosset}, respectively. For WR\,140 and WR\,25, the data were retrieved from the {\it XMM-Newton} archive and processed with the Science Analysis Software version 14.0. The EPIC spectra of each object were fitted between 6 and 8\,keV using a powerlaw with zero slope for the continuum and a single Gaussian for the Fe K line\footnote{The spectral resolution of the EPIC instrument is such that the full Fe {\sc xxv} complex can be represented by a single Gaussian.}. The fluxes of the Fe K line are displayed as a function of orbital phase in Fig.\,\ref{figure8}. In this figure, they are compared against the $1/r$ trend expected for an adiabatic wind-wind interaction zone.

As we can see on this figure, there are clear deviations from the $1/r$ trend. For the shortest period systems, Cyg\,OB2 \#8a and WR\,21a, these deviations are likely due to the shocked gas becoming radiative around periastron passage and/or the shock collapsing onto the surface of the star with the weaker wind \citep{Cazorla,Gosset}. The agreement is better for the longer period systems, although also here there are significant deviations. For instance, in the case of WR\,140, the data follow the relation rather well up to $\Phi = 0.986$, but fall short of the expected emission level afterwards. WR\,140, Cyg\,OB2 \# 8a and Cyg\,OB2 \#9 are prominent non-thermal radio emitters and, in the particular case of WR\,140, \cite{PD} suggested that the deviations from the $1/r$ behaviour for the overall X-ray flux could reflect the impact of particle acceleration on the energy budget and the properties of the shocks \citep[see also the case of 9~Sgr discussed by][]{9Sgr}. Whatever the exact reason for the behaviour seen in Fig.\,\ref{figure8}, it is clear that high-resolution X-ray spectra around the Fe K with next generation observatories will shed new light on the properties of the wind interaction zone.  

\section{Conclusions and future prospects}
In this paper, we have used a rather simple model, based on the analytical solution of \citet{Canto}, to predict the morphology of the Fe {\sc xxv} lines in the X-ray spectra of colliding wind massive binaries in the adiabatic regime. Our results are in qualitative agreement with those of \citet{Henley}. The latter authors computed synthetic line profiles based on the density and velocity fields from snapshots of 2-D axisymmetric hydrodynamical simulations of adiabatic wind collisions and assuming that the winds collide at their terminal velocity. Compared to \citet{Henley}, the advantage of our method is its simplicity and low computational cost, allowing to compute large grids of models for comparison with actual observations. 

We have shown that the morphology of the Fe {\sc xxv} line and its orbital changes provide direct diagnostics of the colliding wind interaction and thus of the properties of the stellar winds. This line offers the cleanest probe of the conditions near the apex of the shock region. On the one hand, it does not suffer from significant absorption by the cool unshocked winds. On the other hand, it is not affected by contributions from the intrinsic emission of the stars that make up the binary system. In this regard, there is a great potential for observational studies of colliding wind binary systems with the bolometric spectrographs that will fly on the coming X-ray observatories {\it Astro-H} and {\it Athena}.

In the future, we will try to generalize this work also to systems where radiative cooling in the shocked winds is efficient and where the Coriolis force leads to a significant aberration. An interesting case of such a system is V\,444~Cyg (WN5 + O6, $P_{\rm orb} = 4.21$\,days) which should be largely in the radiative regime and exhibits a rather prominent Fe {\sc xxv} line in its {\it XMM-Newton} spectra \citep{Lomax}. These properties make V\,444~Cyg an ideal case to apply the Doppler tomography technique that we have outlined above. 



\section*{Acknowledgements}
The Li\`ege team acknowledges support through an ARC grant for Concerted Research Actions, financed by the Federation Wallonia-Brussels, from the Fonds de la Recherche Scientifique (FRS/FNRS), as well as through an XMM PRODEX contract.


\end{document}